\documentclass [a4paper,11pt] {article}

\usepackage {amsmath}

\newcommand {\be}{\begin{equation}}
\newcommand {\ee}{\end{equation}}
\newcommand {\ben}{\begin{eqnarray}}
\newcommand {\een}{\end{eqnarray}}

\begin{document}

\global\parskip=4pt
\titlepage  \noindent
{
   \noindent

\hfill GEF-TH-5/2005  

\vspace{2cm}

\noindent
{\bf
{\large  Instabilities of Noncommutative Two Dimensional BF Model 
}}

\vspace{.5cm}
\hrule

\vspace{2cm}

\noindent
{\bf 
Alberto Blasi, Nicola Maggiore and Michele Montobbio}

\noindent
{\footnotesize {\it
 Dipartimento di Fisica -- Universit\`a di Genova --
via Dodecaneso 33 -- I-16146 Genova -- Italy and INFN, Sezione di
Genova 
} }

\vspace{2cm}
\noindent
{\tt Abstract~:}
The noncommutative extension of two dimensional BF model is
considered. It is shown that the realization of the noncommutative
map via the Groenewold-Moyal star product leads to instabilities of the
action, hence to a non renormalizable theory.
\vfill\noindent
{\footnotesize {\tt Keywords:}
Noncommutative field theory, Algebraic renormalization, BRS quantization}
\newpage

\section{Introduction}

When attempting to define a noncommutative quantum field theory
\cite{Szabo:2001kg} and wishing also to arrive at a formulation which  
allows explicit amplitude computation, one is faced with the problem of choosing a precise form for the non 
commutative product. One of the most popular choices is the Groenewold-Moyal product
\cite{Groenewold:1946kp,Moyal:1949sk}
which is implemented with a simple 
exponential formula and needs the introduction of an antisymmetric constant tensor $\theta^{\mu\nu}$ having the 
dimensions of an inverse mass squared. It is commonly accepted that this procedure leads to a well defined 
noncommutative theory if the commutative model we begin with is sound. We shall show that this is not always the case
by providing a counterexample. We analyze the topological BF model in two spacetime dimensions
\cite{Birmingham:1991ty,Blau:1989bq,Blasi:1992hq}
having in mind that 
a sound noncommutative extension should be based on the functional identities encoding the symmetries, 
on locality and 
power counting, just as it happens in the commutative case. This procedure, in the standard case, leads to the stability
and anomaly analysis i.e. the model is perturbatively renormalizable if the classical action is the most general local 
functional compatible with the above constraints (stability) and the symmetries are not broken by the radiative 
corrections (anomaly) \cite{Piguet:1995er}. 
Since our goal is to provide a counterexample, we concentrate on the stability aspect and only 
at the first order in $\theta^{\mu\nu}$. Accordingly we shall not try to be exhaustive but we will explicitly show 
that the noncommutative extension of the two dimensional BF model based on the Groenewold-Moyal product is unstable.
To fix the notation we briefly recall the functional equations (BRS identity, Landau gauge, ghost equation and vector
supersymmetry) which form a closed algebraic structure and completely define, together with locality and power counting,
the commutative model \cite{Blasi:1992hq}. 
The first step towards a noncommutative definition is then to extend the algebraic constraints 
when $\theta^{\mu\nu}$ is present. Although the choice might not be unique we adopt a ``minimal'' extension for each 
functional operator and conclude that, in order to preserve the algebra, no $\theta^{\mu\nu}$ contribution is allowed. 
Hence the defining equations remain exactly the same we have in the commutative case. Of course this is not so for the
classical action which acquires, at the first order in $\theta^{\mu\nu}$, a local contribution ($X_{\mu\nu}$) with canonical
dimension equal to four and coupled to $\theta^{\mu\nu}$ itself. We then proceed to check that the usual Groenewold-Moyal
extension respects the algebraic constraints, but also show that there is an additional term passing the algebraic 
filter and this term can in no way be generated by the Groenewold-Moyal product.


\section{The classical action and the symmetries}

In the BRS approach in the Landau gauge \cite{Blasi:1992hq}, the classical action of the BF theory over a two dimensional
Euclidean spacetime is: 
\be \label{1}  	  	 	   			 	   	  		  		   				  		 	   
\begin{split}
S_{BF} & = \frac{1}{2} \int d^2x \ \epsilon^{\mu\nu} F_{\mu\nu}^a \ \phi_a
+ \int d^2x \ s \ (\bar{c}_a \ \partial^\mu A_\mu^a) \\
& + \int d^2x \ \left[ \Omega^\mu_a (sA_\mu^a) + L^a (sc_a) + \rho^a (s\phi_a) \right]
\end{split}
\ee
where $\epsilon^{\mu\nu}$ is the completely antisymmetric Levi-Civita tensor ($\epsilon^{12}=+1$).
All fields belong to the adjoint representation of the gauge group $\mathcal{G}$.
In particular $\phi^a$ are scalar fields and $F_{\mu\nu}^a$ is the field strength
\be
F_{\mu\nu}^a = \partial_\mu A_\nu^a - \partial_\nu A_\mu^a + f^{abc} A_\mu^b A_\nu^c
\ee
with $f^{abc}$ completely antisymmetric real structure constants of $\mathcal{G}$. The product rule for the
generators $\tau^a$ of the Lie algebra of $\mathcal{G}$ is
\be
\tau^a \tau^b = \frac{i}{2} \ f^{abc} \tau_c + \frac{1}{2} \ d^{abc} \tau_c \qquad a,b,c=1,\ldots,dim(\mathcal{G})
\ee
so that
\be
[\tau^a,\tau^b] = i f^{abc} \tau_c \quad \{ \tau^a,\tau^b \} = d^{abc} \tau_c \quad 
\ee
with $d^{abc}$ a completely symmetric tensor of rank 3, and $Tr(\tau^a\tau^b) = \delta^{ab}$.
The ghost fields $c^a$ and the antighost fields $\bar{c}^a$ have a Faddeev-Popov charge respectively
equal to $+1$ and $-1$. The external fields $\Omega_\mu^a$, $L^a$ e $\rho^a$ are introduced to take into account
the nonlinearity of the BRS transformations:
\be
\begin{split}
& sA_\mu^a = -(D_\mu c)^a = - (\partial_\mu c^a + f^{abc} A_\mu^b c^c)\\
& s\phi^a = f^{abc} c^b\phi^c \\
& sc^a = \frac{1}{2} \ f^{abc} c^bc^c \\
& s\bar{c}^a = b^a \\
& sb^a = 0
\end{split}
\ee
The fields $b^a$ are the Lagrange multipliers for the gauge condition. With the above definition the $s$ operator
is nilpotent \cite{Piguet:1995er}
\be
s^2=0
\ee
\newpage
The action (\ref{1}) is characterized by the following set of symmetries and constraints: \cite{Blasi:1992hq}
\begin{enumerate}
\item BRS invariance, expressed by the Slavnov-Taylor identity
\be 			   	  				  		  		 			   	  	   		 	
\begin{split}   	  				  		  		 	   			   		  		 		
& {\cal S}(S_{BF}) = \\
& = \int d^2x \left( \frac{\delta S_{BF}}{\delta \Omega_\mu^a} \frac{\delta S_{BF}}{\delta A_\mu^a} + 
\frac{\delta S_{BF}}{\delta \rho^a} \frac{\delta S_{BF}}{\delta \phi^a} +
\frac{\delta S_{BF}}{\delta L^a} \frac{\delta S_{BF}}{\delta c^a} +
b^a \frac{\delta S_{BF}}{\delta \bar{c}^a} \right) = 0
\end{split}
\ee
\item the Landau gauge
\be 
\frac{\delta S_{BF}}{\delta b^a(x)} = \partial^\mu A_\mu^a(x)
\ee
\item the ghost equation of motion \cite{Blasi:1990xz}, which holds true in the Landau gauge
\be 
\int d^2x \ \left( \frac{\delta}{\delta c^a} + f^{abc} \ \bar{c}^b \frac{\delta}{\delta b^c} \right) S_{BF} \equiv
\mathcal{G}^a S_{BF} = \Delta^a
\ee
where
\be
\Delta^a = \int d^2x \ f^{abc} \ (\Omega_\mu^b A_\mu^c - L^bc^c - \rho^b \phi^c)
\ee
\item the antighost equation of motion
\be 
\left( \frac{\delta}{\delta \bar{c}^a(x)} + \partial^\mu \frac{\delta}{\delta \Omega_\mu^a(x)} \right) S_{BF} \equiv
\bar{\mathcal{G}}^a(x) S_{BF} = 0 
\ee
This condition is not independent from the others: it can be derived from the commutator between the Slavnov-Taylor
operator and the gauge condition \cite{Piguet:1995er} 
\item the supersymmetry
\be 
\mathcal{W}_\mu S_{BF} = \Delta_\mu
\ee
where
\ben
\mathcal{W}_\mu = \int d^2x \ \Big( \epsilon_{\mu\nu} \ \rho^a \frac{\delta}{\delta A_\nu^a} 
- \epsilon^{\mu\nu} \ (\Omega_\nu^a + \partial_\nu \bar{c}^a) \frac{\delta}{\delta \phi^a} \nonumber \\
- A_\mu^a \frac{\delta}{\delta c^a} + (\partial_\mu \bar{c}^a) \frac{\delta}{\delta b^a}
- L^a \frac{\delta}{\delta \Omega_\mu^a} \Big)
\een
\be 
\Delta_\mu = \int d^2x \ \left[ L^a(\partial_\mu c^a) - \rho^a \partial_\mu \phi^a - \Omega_\nu^a(\partial_\mu A_\nu^a)
- \epsilon_{\mu\nu} \rho^a \partial^\nu b^a \right]
\ee
the existence of which is due to the topological nature of the BF model and to the choice of the Landau gauge 
\cite{Maggiore:1991aa}
\end{enumerate}
We also note that the breakings $\Delta^a$ and $\Delta_\mu$, being linear in the quantum fields, will be present
only at the classical level \cite{Piguet:1995er} . \\

The whole set of all these symmetries can be summarized in a closed algebra with breakings, that, for a generic
even Faddeev-Popov charged functional $\gamma$, can be expressed as:
\be \label{2}				 	 		   				 	   		  	 	   			 	 		 
\begin{split}
& B_\gamma {\cal S}(\gamma)=0 \\ 
& \mathcal{W}_\mu {\cal S}(\gamma) + B_\gamma (\mathcal{W}_\mu\gamma - \Delta_\mu) = P_\mu\gamma \\
& \mathcal{G}^a{\cal S}(\gamma)+B_\gamma(\mathcal{G}^a\gamma - \Delta^a) = \mathcal{H}^a\gamma \\
& \mathcal{G}^a(\mathcal{W}_\mu \gamma - \Delta_\mu) + \mathcal{W}_\mu(\mathcal{G}^a\gamma - \Delta^a) = 0 \\
& B_\gamma \left( \frac{\delta\gamma}{\delta b^a}-\partial^\mu A_\mu^a \right) 
- \frac{\delta}{\delta b^a} \ {\cal S}(\gamma) = \bar{\mathcal{G}}^a \gamma\\
& \{ \mathcal{G}^a,\mathcal{G}^b \} =0 \\
& \{ \mathcal{W}_\mu,\mathcal{W}_\nu \} =0 \\
& [\mathcal{H}^a,\mathcal{G}^b] = - f^{abc} \mathcal{G}_c \\
& [\mathcal{H}^a,\mathcal{H}^b] = - f^{abc} \mathcal{H}_c \\
\end{split}
\ee
where $B_\gamma$ is the linearized Slavnov-Taylor operator 
\be
\begin{split}
B_\gamma = \int d^2x \ & \Bigg[ \frac{\delta\gamma}{\delta\Omega_\mu^a}\frac{\delta}{\delta A_\mu^a}
+ \frac{\delta\gamma}{\delta A_\mu^a} \frac{\delta}{\delta\Omega_\mu^a}
+ \frac{\delta\gamma}{\delta\rho^a}\frac{\delta}{\delta\phi^a} 
+ \frac{\delta\gamma}{\delta\phi^a}\frac{\delta}{\delta\rho^a} \\
& + \frac{\delta\gamma}{\delta L^a}\frac{\delta}{\delta c^a}
+ \frac{\delta\gamma}{\delta c^a}\frac{\delta}{\delta L^a}
+ b^a \frac{\delta}{\delta\bar{c}^a} \Bigg]
\end{split}
\ee
the operator $\mathcal{H}^a$ expresses a global gauge transformation
\be
\mathcal{H}^a = \sum_{(all \ fields \ \psi)} \int d^2x \ f^{abc} \ \psi^b \frac{\delta}{\delta \psi^c}
\ee
and finally $P_\mu$ is a global translation  
\be
P_\mu = \sum_{(all \ fields \ \psi)} \int d^2x \ (\partial_\mu \psi^a ) \frac{\delta}{\delta \psi_a}
\ee 
As shown in \cite{Blasi:1992hq}, the symmetries in (\ref{2}) allow a full quantum extension of the theory. 
In fact it can be proved
that the action in (\ref{1}) is stable and
that the symmetries are not anomalous.
\footnote{In effect this theory can be proved to be perturbatively finite}


\section{Noncommutative extension by means of the \\ Groenewold-Moyal product}

In order to extend the classical action (\ref{1}) to a noncommutative spacetime we have to define a new prescription
for the product of fields. A popular choice in the literature is the Groenewold-Moyal star product defined by 
\cite{Groenewold:1946kp,Moyal:1949sk}
\be
\begin{split}
f(x) \star g(x) &= f(x) \ e^{\frac{i}{2} \overleftarrow{\partial_i}\theta^{ij} \overrightarrow{\partial_j}} \ g(x) \\
&=f(x)g(x) + \frac{i}{2} \ \theta^{ij} \partial_i f(x) \partial_j g(x) + O(\theta^2)
\end{split}
\ee
where $\theta^{ij}$ is an antisymmetric real constant tensor, with the dimensions of an inverse squared mass, defined by 
\be
[\hat{x}^i,\hat{x}^j] = i \ \theta^{ij}
\ee 
At the leading order in $\theta$, we find that the noncommutative
extension of the classical action (\ref{1}) according to the
Groenewold-Moyal star product is
\be
\widehat{S}_{BF} = S_{BF} + \theta^{\rho\sigma} X^{(GM)}_{\rho\sigma} + O(\theta^2)
\ee
where
\be \label{4}		 		   	  		 						  			 		  		   	   
\begin{split}
& X^{(GM)}_{\rho\sigma} = \frac{1}{4} \int d^2x \ d^{abc} \epsilon^{\mu\nu} 
(\partial_\sigma A_\nu^a)(\partial_\rho A_\mu^b) \ \phi^c \\
& + \frac{1}{2}\int d^2x \ d^{abc} \  (\partial_\mu \bar{c}^a)(\partial_\rho c^b)(\partial_\sigma A_\mu^c)
+ \frac{1}{2}\int d^2x \ d^{abc} \ \Omega_\mu^a (\partial_\rho c^b)(\partial_\sigma A_\mu^c) \\
& + \frac{1}{4} \int d^2x \ d^{abc} \ L^a (\partial_\rho c^b)(\partial_\sigma c^c) 
+ \frac{1}{2}\int d^2x \ d^{abc} \ \rho^a(\partial_\rho c^b)(\partial_\sigma \phi^c)
\end{split}
\ee


\section{Symmetries of the noncommutative action}

Generally speaking, we may say that the noncommutative extension of our classical action $S_{BF}$ has the
following form:
\be \label{3}					   		 		 	 			  	  			  			   		  
\widehat{S}_{BF} = S_{BF} + \theta^{\rho\sigma} X_{\rho\sigma} + O(\theta^2)
\ee
where $X_{\rho\sigma}$ is a local functional of the fields, with canonical dimension less than or equal to four
($[\theta^{\rho\sigma}]=-2$) and Faddeev-Popov charge zero. In the previous section we have found a
particular extension $X^{(GM)}_{\rho\sigma}$, which derives from the use of the Groenewold-Moyal product. 
This approach is neither the most general, nor the unique one. In this paper we want to present an alternative way 
of constructing the correction $X_{\rho\sigma}$ without introducing any particular star product a priori defined. 
Our point of view, which is borrowed from the standard approach to the commutative theory, is that the symmetries
themselves, once correctly defined to take into account the $\theta^{\rho\sigma}$ tensor, should be used as guidelines
to characterize the $X_{\rho\sigma}$ term. The main ingredient is thus the algebra in (\ref{2}), which we want
to preserve. Of course the definition of the symmetries in (\ref{2}) when $\theta^{\rho\sigma}$ is present has a 
certain degree of arbitrariness. Our choice is ``minimal'', in the sense that we decide to keep unchanged the 
functional form of the operators that express the symmetries and constraints on $S_{BF}$, and the introduction 
in (\ref{3}) of the tensor $\theta^{\rho\sigma}$ can only modify the classical breakings
\be \label{6}					  				  				  	   		   	   		 				
\begin{split}
& \frac{\delta \widehat{S}_{BF}}{\delta b^a(x)} = \partial^\mu A_\mu^a(x) + \theta^{\mu\nu} \Xi_{\mu\nu}^a(x) + O(\theta^2) \\
& \mathcal{G}^a \widehat{S}_{BF} = \Delta^a + \theta^{\mu\nu} \int d^2x \ \Delta_{\mu\nu}^a(x) + O(\theta^2) \\
& \mathcal{W}_\mu \widehat{S}_{BF} 
= \Delta_\mu + \theta^{\rho\sigma} \int d^2x \ (\Lambda_\mu)_{\rho\sigma}(x) + \theta^{\mu\tau} \int d^2x \ \Lambda_\tau(x) + O(\theta^2)
\end{split}
\ee
A detailed analysis of the first order terms in $\theta^{\mu\nu}$ in (\ref{6}) shows that none of them is allowed, if we wish
to preserve the algebraic structure in (\ref{2}). This can be proved by explicitly evaluating the commutators and the 
anticommutators in (\ref{2}) with the objects defined in (\ref{6}) and enforcing the validity of (\ref{2}). We may now assume that all the symmetries 
characterizing $S_{BF}$ can be extended to $\widehat{S}_{BF}$. This is of great relevance because it expresses a 
strong constraint on the choice of the possible corrections $X_{\rho\sigma}$ to the ordinary action.


\section{Comparison with the Groenewold-Moyal extension}

A direct calculation shows that the noncommutative corrections derived with the use of the Groenewold-Moyal star product are all 
compatible with the symmetries characterizing $\widehat{S}_{BF}$. However they are not the only ones. For example a term of the form
\be \label{5} 		  			 	  		   			  			  	  							
\theta^{\rho\sigma} \int d^2x \ F_{\rho\mu}^a \epsilon^{\mu\nu} F_{\nu\sigma}^a
\ee
is compatible with the constraints, but there is no way to derive it from the Groenewold-Moyal product. The presence of a term of the form
(\ref{5}), that is compatible with all of the algebraic constraints, implies that the noncommutative extensions of the BF model 
based on the Groenewold-Moyal star product are not stable. Consequently their quantum extensions cannot be correctly defined since the coupling
(\ref{5}), needed as a counterterm, is not present at the classical level
\cite{Piguet:1995er}. On the other hand, if we add (\ref{5}) to the classical
action we definitely spoil the Groenewold-Moyal star product extension.


\section{Conclusions}

In this letter we have shown that the noncommutative extension of the two dimensional BF model based on the Groenewold-Moyal
product is not stable in the sense that, already at the first order in $\theta^{\mu\nu}$, the resulting classical action
is not the most general local functional respecting power counting and satisfying the algebraic constraints which define
the model. Two final remarks are in order: first the term in (\ref{5}) can be rewritten (due to the two spacetime 
dimensions) as 
\be
\theta^{12} \ \int d^2x \ F_{\mu\nu}^a \ F^{\mu\nu}_a
\ee
which shows that the noncommutative stable extension acquires a contribution which in two spacetime dimensions 
has still a  topological character, but in a more general context is not topological, suggesting that some of the
symmetries could be broken at the quantum level. Second, the same term in (\ref{5}) will propagate to all orders in 
$\theta^{\mu\nu}$, by simply taking powers of $\theta^{\mu\nu} X_{\mu\nu}$ and mixes with the Groenewold-Moyal contributions.
This suggests that, at an arbitrary order in $\theta^{\mu\nu}$, we will find non
Groenewold-Moyal terms whose coefficient is free 
and thus an infinite number of couplings. In our opinion this opens a serious problem concerning the noncommutative 
extensions of quantum field theory models. Work is in progress to completely characterize the BF model. The analysis 
will include not only stability but also the anomaly issue. In particular the possible presence of anomalous terms will
definitely spoil the renormalizability of the noncommutative extension, contrary to the standard commutative case.



\begin{thebibliography}{20}

\bibitem{Szabo:2001kg}
  R.~J.~Szabo,
  {\it Quantum field theory on noncommutative spaces},
  Phys.\ Rept.\  {\bf 378}, 207 (2003)
  [arXiv:hep-th/0109162].
\bibitem{Groenewold:1946kp}
  H.~J.~Groenewold,
  {\it On The Principles Of Elementary Quantum Mechanics},
  Physica {\bf 12}, 405 (1946).
\bibitem{Moyal:1949sk}
  J.~E.~Moyal,
  {\it Quantum Mechanics As A Statistical Theory},
  Proc.\ Cambridge Phil.\ Soc.\  {\bf 45}, 99 (1949).
\bibitem{Birmingham:1991ty}
  D.~Birmingham, M.~Blau, M.~Rakowski and G.~Thompson,
  {\it Topological field theory},
  Phys.\ Rept.\  {\bf 209}, 129 (1991).
\bibitem{Blau:1989bq}
  M.~Blau and G.~Thompson,
  {\it Topological Gauge Theories Of Antisymmetric Tensor Fields},
  Annals Phys.\  {\bf 205}, 130 (1991).
\bibitem{Blasi:1992hq}
  A.~Blasi and N.~Maggiore,
  {\it Infrared and ultraviolet finiteness of topological BF theory in
  two-dimensions},
  Class.\ Quant.\ Grav.\  {\bf 10}, 37 (1993)
  [arXiv:hep-th/9207008].
\bibitem{Piguet:1995er}
  O.~Piguet and S.~P.~Sorella,
  {\it Algebraic renormalization: Perturbative renormalization, symmetries and
  anomalies},
  Lect.\ Notes Phys.\  {\bf M28}, 1 (1995).
\bibitem{Blasi:1990xz}
  A.~Blasi, O.~Piguet and S.~P.~Sorella,
  {\it Landau Gauge And Finiteness},
  Nucl.\ Phys.\ B {\bf 356}, 154 (1991).
\bibitem{Maggiore:1991aa}
  N.~Maggiore and S.~P.~Sorella,
  {\it Finiteness of the topological models in the Landau gauge},
  Nucl.\ Phys.\ B {\bf 377}, 236 (1992).
\end{thebibliography}
\end{document}